\documentclass[twocolumn,superscriptaddress,showpacs,amsmath,amssymb,pra,longbibliography]{revtex4-1}

\usepackage[pdftex]{graphicx} 
\usepackage{dcolumn}  
\usepackage{bm}       
\usepackage[usenames,dvipsnames]{color}
\definecolor{URLCOL}{rgb}{0,0.52,0.83} 
\definecolor{LINKCOL}{rgb}{0.05,0.5,0} 
\definecolor{orange}{rgb}{0.6,0.3,0} 
\definecolor{CITECOL}{rgb}{0.25,0,0.48} 
\usepackage{epstopdf}

\usepackage[encapsulated]{CJK}

\usepackage{hyperref}

\usepackage{booktabs}
\usepackage{natbib}

\usepackage{tabulary}
\usepackage{ulem}

\definecolor{TITLECOL}{rgb}{0.1,0.2,0.7} 
\definecolor{SECOL}{rgb}{0.1,0.2,0.7} 
\definecolor{CONTENTSCOL}{rgb}{0.1,0.2,0.7} 
\definecolor{SSECOL}{rgb}{0.25,0,0.48} 
\definecolor{SSSECOL}{rgb}{0.2,0.08,0.53} 
\definecolor{FINCOL}{rgb}{0.01,0.3,0.07} 

\definecolor{URLCOL}{rgb}{0,0.17,0.43} 
\definecolor{LINKCOL}{rgb}{0.05,0.4,0} 
\definecolor{CITECOL}{rgb}{0.35,0,0.48} 

\def\sss{\scriptscriptstyle\rm}

\def\bea{\begin{eqnarray}}
\def\eea{\end{eqnarray}}
\def\ben{\begin{equation}}
\def\een{\end{equation}}
\def\benu{\begin{enumerate}}
\def\enu{\end{enumerate}}

\def\bei{\begin{itemize}}
\def\eei{\end{itemize}}
\def\beit{\begin{itemize}}
\def\eit{\end{itemize}}
\def\benu{\begin{enumerate}}
\def\enu{\end{enumerate}}

\def\br{{\bf r}}

\def\s{_{\sss S}}

\def\xc{_{\sss XC}}

\def\ee{_{\rm ee}}

\def\ML{^{\rm ML}}

\def\n{n}

\def\Tabref#1{Table \ref{#1}}

\def\Figref#1{Fig.\ \ref{#1}}

\begin{document}
\begin{CJK*}{UTF8}{gbsn}
\title{
Pure density functional for strong correlations and the thermodynamic 
limit from machine learning
}
\author{Li Li (李力)}
\affiliation{Department of Physics and Astronomy, University of California, Irvine, CA 92697}
\author{Thomas E. Baker}
\affiliation{Department of Physics and Astronomy, University of California, Irvine, CA 92697}
\author{Steven R. White}
\affiliation{Department of Physics and Astronomy, University of California, Irvine, CA 92697}
\author{Kieron Burke}
\affiliation{Department of Chemistry, University of California, Irvine, CA 92697}
\affiliation{Department of Physics and Astronomy, University of California, Irvine, CA 92697}
\date{\today}

\begin{abstract}
We use density-matrix renormalization group, applied to a one-dimensional
model of continuum Hamiltonians, to accurately solve chains of hydrogen
atoms of various separations and numbers of atoms.  We train and
test a machine-learned approximation to
$F[\n]$, the universal part of the electronic density
functional, to within quantum chemical accuracy.
Our calculation (a) bypasses the standard Kohn-Sham approach,  avoiding the
need to find orbitals, (b) includes
the strong correlation of highly-stretched bonds without any specific
difficulty (unlike all standard DFT approximations) and (c) is so accurate
that it can be used to find the energy in the thermodynamic limit
to quantum chemical accuracy.
\end{abstract}

\maketitle
\end{CJK*}

\section{Introduction}

Although widely used in solid-state physics, chemistry, and materials
science \cite{PGB14},
Kohn-Sham density functional theory (KS-DFT) with standard approximations
fails for strong correlation \cite{KS65,WBSB14}.  The prototype is the H$_2$ molecule.
When stretched, the electrons localize on each site while
remaining in a singlet state, but this is not captured by such calculations \cite{PSB95}.
These kinds of difficulties have led to the popularity of many 
beyond-DFT schemes, ranging from
the simple addition \cite{LAZ95} of a Hubbard $U$ to the use of dynamical mean
field theory \cite{GKKR96} as well as many others.  

But even KS-DFT is too slow for many large calculations, such as those
using classical MD or continuum mechanics \cite{IMT05}.  The original DFT, first
suggested in the Thomas-Fermi approximation \cite{T27,F27} and later justified by the
Hohenberg-Kohn theorem \cite{HK64}, uses only pure functionals of the total density, $\n(\br)$.
This `orbital-free' version has the potential to be much faster than
even the most efficient KS implementations, because the KS equations need
not be solved \cite{TWb02}.  Several recent attempts have
constructed machine learning (ML)
kinetic energy functionals 
specifically to bypass this step \cite{SRHM12,SRHB13,LSPH14,VSLR15}.  These are designed to be used
in conjunction with standard KS approximations to speed up such calculations,
but not to improve their accuracy. 

Meanwhile, beyond the world of DFT, density matrix renormalization group (DMRG)
has become a standard tool for finding extremely accurate solutions to 
strongly correlated lattice problems \cite{S91,S92,U05,U11}.  In recent years, a one-dimensional
analog of ab-initio Hamiltonians has been developed, using typically about
20 grid points per atom and interactions involving many grid points, with the
express purpose of rapidly exploring both conceptual and practical issues in
DFT \cite{SWWB12,WSBW12,WSBW13,WBSB14,BSWB15}.
A particular advantage is that, since 2000 grid points is routinely 
accessible, this includes up to 100 atoms, and extrapolations to the thermodynamic
limit are much easier than in three dimensions.  Applications include a demonstration
of the behavior of the KS gap in a Mott-Hubbard insulator \cite{SWWB12} and a proof of
convergence of the KS equations with the exact functional, regardless of the
starting point or strength of correlation \cite{WSBW12}.

\begin{figure}[htb]
\includegraphics[width=0.8\columnwidth]{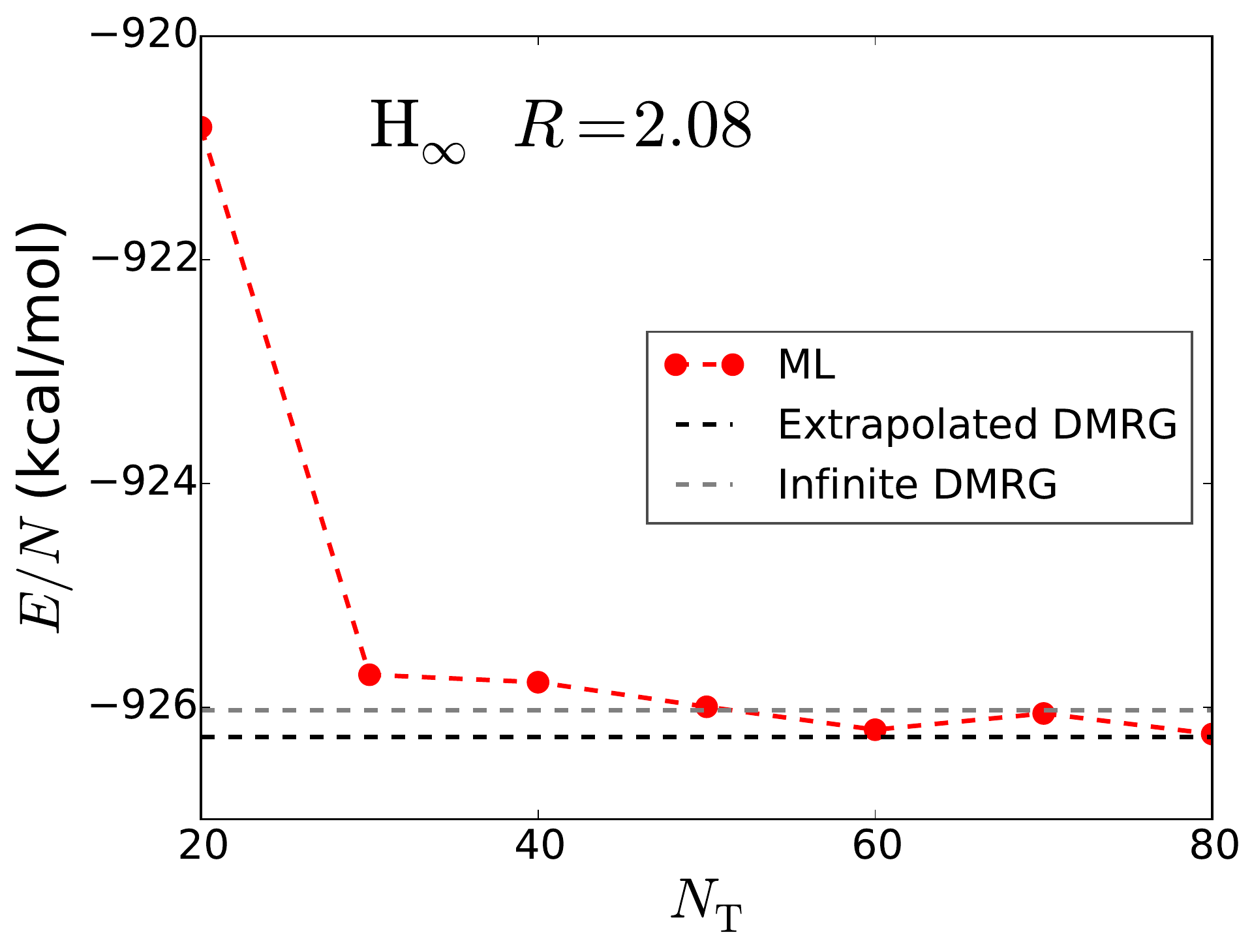}
\caption{(Color online) Electronic energy of infinite chain from
model learned from extrapolated chain densities and energies.  The
accurate value was calculated with infinite DMRG (see text).}
~
\vskip -0.8cm
\label{Einf}
\end{figure}
In the present work, we combine all these methodologies to demonstrate several
important features.  We perform DMRG calculations on a variety of one-dimensional
hydrogen atom chains, with from 2 to 20 atoms, and whose interatomic
spacing $R$ varies from 1 to 10 Bohr radii, and use
these to train a ML model of $F[\n]$, the `universal' part of the
density functional identified by Hohenberg-Kohn.  This simulanteously includes
both the non-interacting kinetic energy sought in orbital-free DFT and the 
exchange-correlation energy that is approximated in KS calculations.  We demonstrate
that, with reasonable amounts of training, we can {\em self-consistently} calculate
densities and energies for these chains at new values of $R$, outside the 
training set, with quantum chemical accuracy.   
This includes highly stretched systems which are strongly correlated, and where
all popular DFT approximations fail.
We furthermore extrapolate the
DMRG densities from the center of finite chains to the infinite chain limit,
i.e., a 1d solid.  We train a new ML model  and find we can solve
self-consistently the solid problem at the same level of accuracy.  
Fig. \ref{Einf} shows the convergence of our ML method for a typical separation
of the infinite chain with respect to the number of training points.
The horizontal lines show two independent DMRG estimates of the energy.

\section{Background}

\subsection{DFT}
The Hohenberg-Kohn theorem \cite{HK64} establishes that the ground-state
energy and density of a many-body problem may be found by
minimizing a density functional:
\ben
E = \min_\n \left\{ F[\n] + \int d^3r\, \n(\br)\, v(\br)\right\},
\een
where $\n(\br)$ is the single-particle density, normalized to $N$
particles, and $v(\br)$ is the one-body potential.  The functional
$F$ can be defined via a constrained search as \cite{L79}
\ben
F[\n]=\min_{\Psi\to\n} \langle \Psi |\, \hat T + \hat V\ee | \Psi \rangle
\een
where $\hat T$ is the kinetic energy operator and $V\ee$ is the
electron-electron repulsion operator, while $\Psi$ is a normalized
antisymmetric (for fermions) wavefunction.  While this showed that
the old Thomas-Fermi theory \cite{T27,F27,LS77} was an approximation to an exact formulation,
few modern calculations perform such a direct minimization.  In practice,
almost all calculations use the famous Kohn-Sham (KS) scheme, which uses
an auxillary set of non-interacting orbitals in a single, multiplicative
potential whose density is defined to match that of the original system,
and in terms of which we can write
\ben
F[\n] = T\s[\n] + U [\n] + E\xc[\n],
\een
where $T\s$ is the non-interacting kinetic energy of the KS electrons,
$U$ is the Hartree self-repulsion, and $E\xc$ is the exchange-correlation
energy (defined by this equation).  

The genius of the KS formulation is that
$E\xc$ is typically a small fraction of $F$, so that much higher accuracy
can be achieved by approximating only this component.
The cost of the KS scheme is formally $N^3$, the cost of solving for the
orbitals.
Much of modern DFT research is devoted to improving approximations to
$E\xc$, within which all quantum-many body effects are contained (by definition).
The smaller field of pure DFT, also known as `orbital-free', aspires to
approximate $T\s[\n]$ directly, as in the old TF theory \cite{T27,F28}, and thus bypass the
need to solve the KS equations. 

\begin{figure}[htb]
\includegraphics[width=0.8\columnwidth]{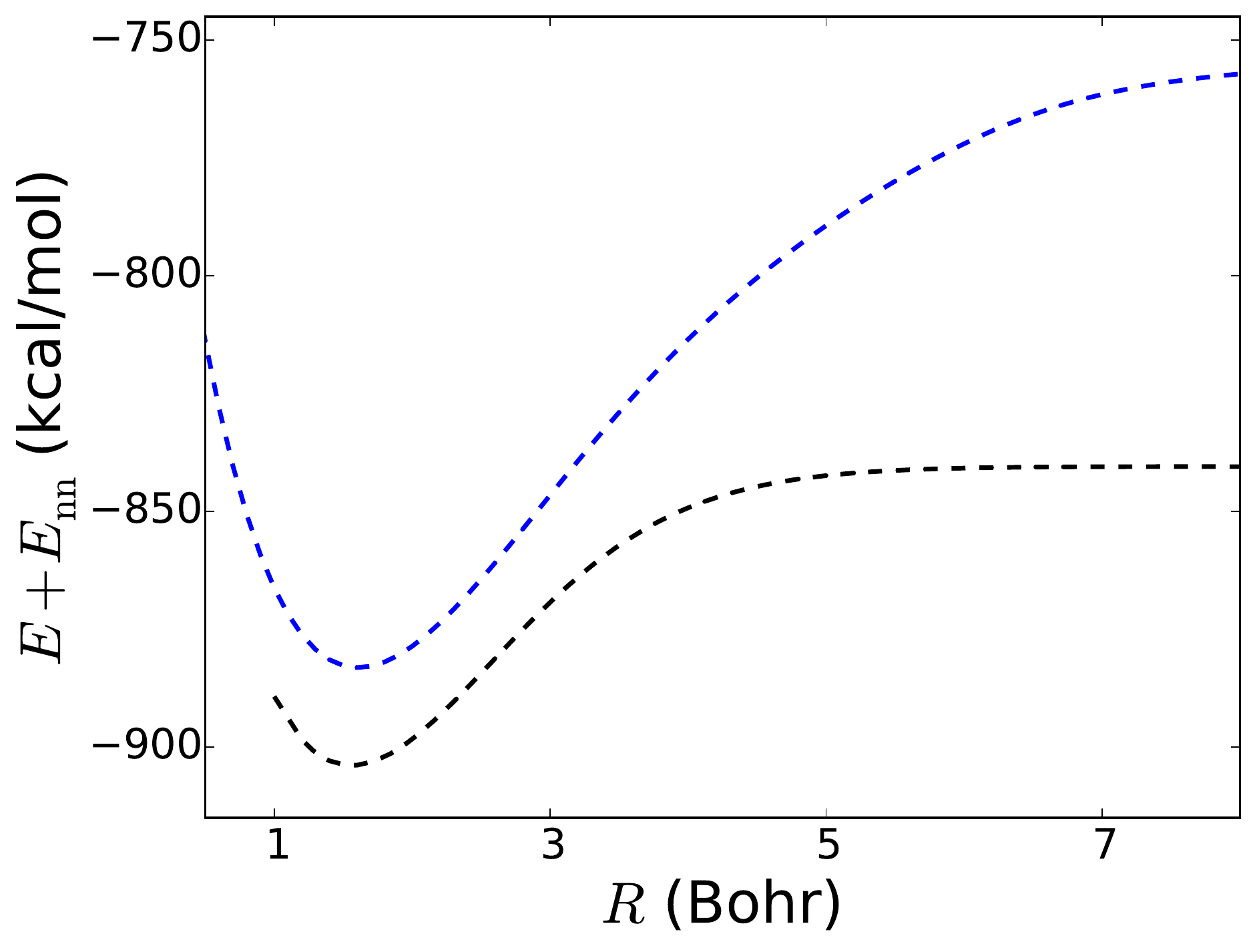}
\caption{(Color online) Binding curve for a 1d H$_2$ molecule.  Black: highly accurate, converged
DMRG results.  Blue: LDA result restricted to a spin singlet \cite{BSWB15}.}
~
\vskip -0.8cm
\label{h2}
\end{figure}
Many modern XC approximations are local or semilocal, i.e.,
use the density and its gradient to approximate the XC
energy density at a point.  While remarkably useful results
can be obtained with such approximations, there remains a
classic failure that can be understood in terms of the simple
H$_2$ molecule \cite{CMY08}.  Those approximations work well in the vicinity
of the equilibrium bond length, but as the bond is stretched, they
fail more and more badly.  In the limit of a large but finite
bond length (which we call stretched), a spin-restricted calculation
yields the highly inaccurate energy of two unpolarized H atoms.
On the other hand, an unrestricted calculation yields an accurate
stretched energy, but has broken spin symmetry.  Neither situation
is satisfactory \cite{PSB95}, and most modern approximations fail in this way.
An analogous failure occurs for semilocal approximations to $T\s[\n]$
when bonds are stretched in orbital-free DFT.   
Fig.~\ref{h2} illustrates the failure of semilocal XC,
by comparing the blue restricted LDA curve with the
black DMRG curve.  There is a huge error in the stretched limit.

\subsection{DMRG benchmark data}

It is difficult to overemphasize the utility of benchmark quantum
chemical calculations for the development of DFT.  The 
DFT revolution in quantum chemistry was made possible by the
existence of the well-tested G2 data set for small molecules, 
and databases in quantum chemistry have proliferated ever since.
On the other hand, calculations of `quantum chemical' accuracy,
i.e., errors below 1 kcal/mol,
are much more difficult and rarer for solids.  A recent heroic 
effort \cite{YHUM14} was made for benzene, a molecular crystal.

For the present study, we need to consider chains of up to 20 H atoms, with many different values of the interatomic
spacing ranging from about 1 to 10 Bohr.  Extracting this large
amount of data at the required level of accuracy from a quantum
chemical code would be extremely demanding, if not impossible, given
the strong correlation effects when the bonds are stretched.

Recently, DMRG has been applied to a one-dimensional analog of 
real-space Coulomb-interacing Hamiltonians, for precisely the
purpose of performing demanding, highly accurate benchmark
calculations of strongly correlated systems.  In particular, the
interaction is modeled as 
\ben
v\ee(u) = A\, \exp(-\kappa\, |u|)
\een 
where $A=1.071$ and $\kappa^{-1}=2.385$ \cite{BSWB15}.
This choice best mimics a popular
soft Coulomb interaction, while having a single exponential allows
DMRG to run very fast \cite{BSWB15}.  The one-body potential is then taken as
$v(x)=-Z v\ee(x)$, where $Z$ is the `charge' on a nucleus. Here
$Z=1$ for each H atom in the chain.  This 1d analog allows rapid
testing of novel ideas in electronic structure, especially those
involving the bulk limit.  Fig. \ref{h2} is in fact for 1d H$_2$
with these parameters, and illustrates that the failures of standard
DFT approximations such as LDA mimic those of 3d Coulomb systems.
The DMRG curve plateau is at twice the ground-state energy of one of these
1d H atoms.

\subsection{Machine learning of the KS kinetic energy functional}
\def\NT{N_\mathrm{T}}

ML is a set of algorithms developed to find hidden insights in data.
It is widely used especially when the pattern behind complicated data is
difficult to deduce explicitly. 
Successful applications include computer vision \cite{PVJZ11},
cybersecurity \cite{ZLW16}, ancient abstract strategy games \cite{SHMG16}, etc.

Recently, in chemistry and materials science,
machine-learning has become a popular tool for analyzing
properties of molecules and materials, and finding specific functions
from large data sets \cite{RTML12, R16}.  But it has also been applied to the 
problem of finding density functionals, constructed by interpolation from 
accurate examples.   To date, the focus has been on the KS kinetic
energy, $T\s[\n]$, partially because of the ready availability of
data (every cycle of every one of the 30,000 KS-DFT calculations each
year \cite{PGB14} produces an accurate example of $T\s[\n]$) and because of the enormous
potential for speeding up routine DFT calculations.

The ML algorithm we used for modeling $T\s[n]$ is kernel ridge regression (KRR). 
It is a nonlinear regression method with an L2 regularization \cite{HTFF05}. 
The density functional is represented as
\ben
T\s\ML[n] = \sum^{\NT}_{i=1}\alpha_i k[n,n_i],
\een
where $\NT$ is the number of training data, 
$n_i(x)$ are the training data and $k[n,n_i]$ is a kernel, some measure of the
``similarity'' between densities.  Throughout this work, we use a Gaussian kernel,
\ben
k[n,n']=\exp (-\| n - n' \|^2/2\sigma^2),
\een
where 
\ben
\|n-n'\| = \int dx (\n(x)-\n'(x))^2.
\een
Such a kernel is standard in KRR, and has yielded excellent results in
previous studies of $T\s[\n]$ \cite{LSPH14}.
The weights ${\bf \alpha}=(\alpha_1, \cdots, \alpha_{N_\mathrm{T}})$ are found by optimizing
the cost function
\ben
\mathcal{C}({\bf \alpha}) = 
\sum^{N_\mathrm{T}}_{i=1}(T\s\ML[n_i] - T\s[n_i])^2 + \lambda {\bf \alpha}^\mathrm{T}{\bf K}{\bf \alpha}
\een
where ${\bf K}$ is the kernel matrix, $K_{ij}=k[n_i,n_j]$. 
The regularization strength $\lambda$ and length scale $\sigma$ are
hyperparameters which are found via cross validation. 
A crucial principle in kernel ridge regression is the separation of the training data
from the test data.  A test set is constructed entirely independently from the
training set.  The cross-validation to find the hyperparameters occurs using only
training data.  The resulting approximate functional is tested only on the test data.

While highly accurate results for $T\s[\n]$ can be found with relatively little
data \cite{SRHM12}, it was immediately realized that the corresponding functional derivative
is highly {\em inaccurate}.   This is unfortunate, as the practical usefulness of an accurate
model for 
$T\s[\n]$ is in finding the density via solution of the Euler equation (for the KS
system): 
\ben
\frac{\delta T\s}{\delta \n(x)}= - v\s(x),
\een
where $v\s(x)$ is the KS potential. This difficulty has been surmounted in
a sequence of increasingly sophisticated methods \cite{SRHB13,LSPH14,VSLR15},
each of which constrains the
density search to only the manifold of densities spanned by the data, avoiding
searching in directions for which there is insufficient data to evaluate $T\s$
accurately.  With such techniques, it has been possible to demonstrate an ML
$T\s$ functional that correctly mimics the KS solution even as a bond stretches \cite{SRHB13},
something impossible for any local or semilocal approximation to $T\s$.  
The value of this is to cut down the computational cost of large, repetitive 
KS calculations, but one still uses some standard XC approximation.
Thus a machine-learned functional for $T\s$ can, at best, reproduce the incorrect
LDA curve of Fig. \ref{h2}.

\section{Method}

In all applications in this work, we generate a large data set of highly
accurate results generated using DMRG.  We use a real-space grid with
spacing 0.04,
which has previously been shown to be sufficient to converge the results \cite{BSWB15}.
We calculate the energies and densities of chains of even numbers of atoms, from 2 to 20,
with interatomic separations between 1 and 10 Bohr. Higher accuracy is achieved when
every atom is centered on a grid point, discretizing the set of allowed separations.
The specific separations used 
are listed in the supplemental information.

Then a subset of these data are left out as test set. The training set, with $\NT$ values of $R$,
are collected from the remaining data. These are chosen to be as close to equally spaced as practical.
The test set is excluded from the data where the training set is sampled from, to avoid contamination via the cross-validation
process.

\subsection{Machine-learned functional for a given molecule}

\begin{figure}[htb]
\includegraphics[width=0.8\columnwidth]{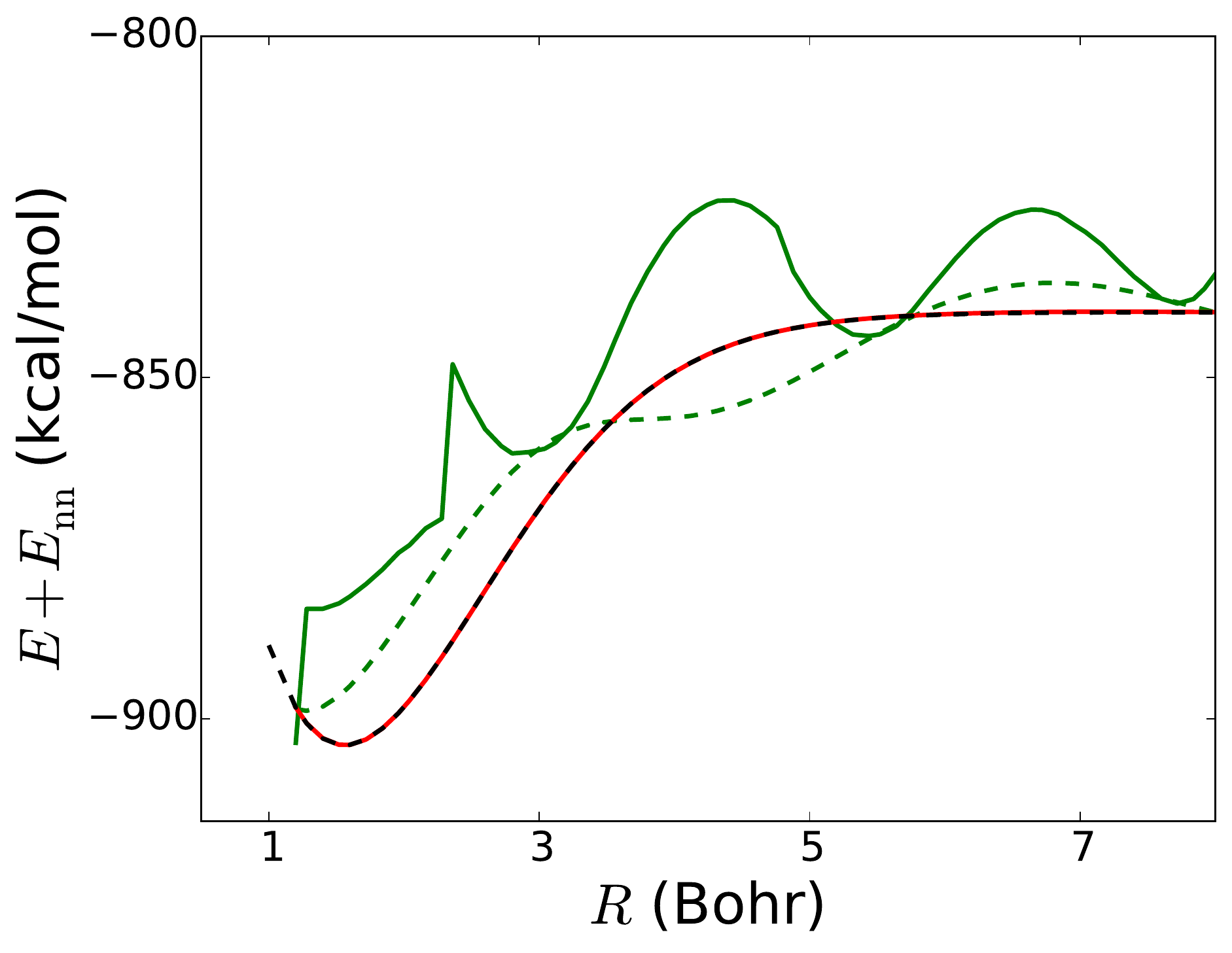}
\caption{(Color online) Same as Fig. \ref{h2}. The green curves are ML with $\NT=5$
on both the exact (dashed) and ML-optimized (solid) densities. The red solid curve is the
ML with $N_T$=20 on ML-optimized (solid) densities. Black dashed curve is the exact DMRG curve, matching nearly exactly the $N_T$=20 on ML line.
}
~
\vskip -0.8cm
\label{h2a}
\end{figure}

We continue to use the H$_2$ molecule to illustrate our method.
Contrary to previous work, we apply KRR algorithms to ML the
interacting functional $F[\n]$ itself, by training on highly
accurate DMRG energies and densities at various values of
$R$.
In \Tabref{t:results}, we list the errors for H$_2$ as a function of $\NT$,
both on the exact density and on an optimally constrained density
found by the methods of Ref.~\cite{SRMB15}.  

To illustrate the procedure, in
Fig. \ref{h2a}, we show the energies with only 5 training points,
$R=1.00, 3.20, 5.48, 7.76, 10.00$, yielding the smooth, green dashed curve, when evaluated on
the exact densities.  The curve (almost) exactly matches at the training points, but is
noticeably inaccurate inbetween.  But note that, in contrast to all previous studies, we
are fitting the full $F[\n]$, not just $T\s[\n]$, so that, e.g., our inaccurate curve
dissociates H$_2$ correctly, while no standard DFT calculation, with a standard XC
approximation, can.

\begin{figure}[htb]
\includegraphics[width=0.8\columnwidth]{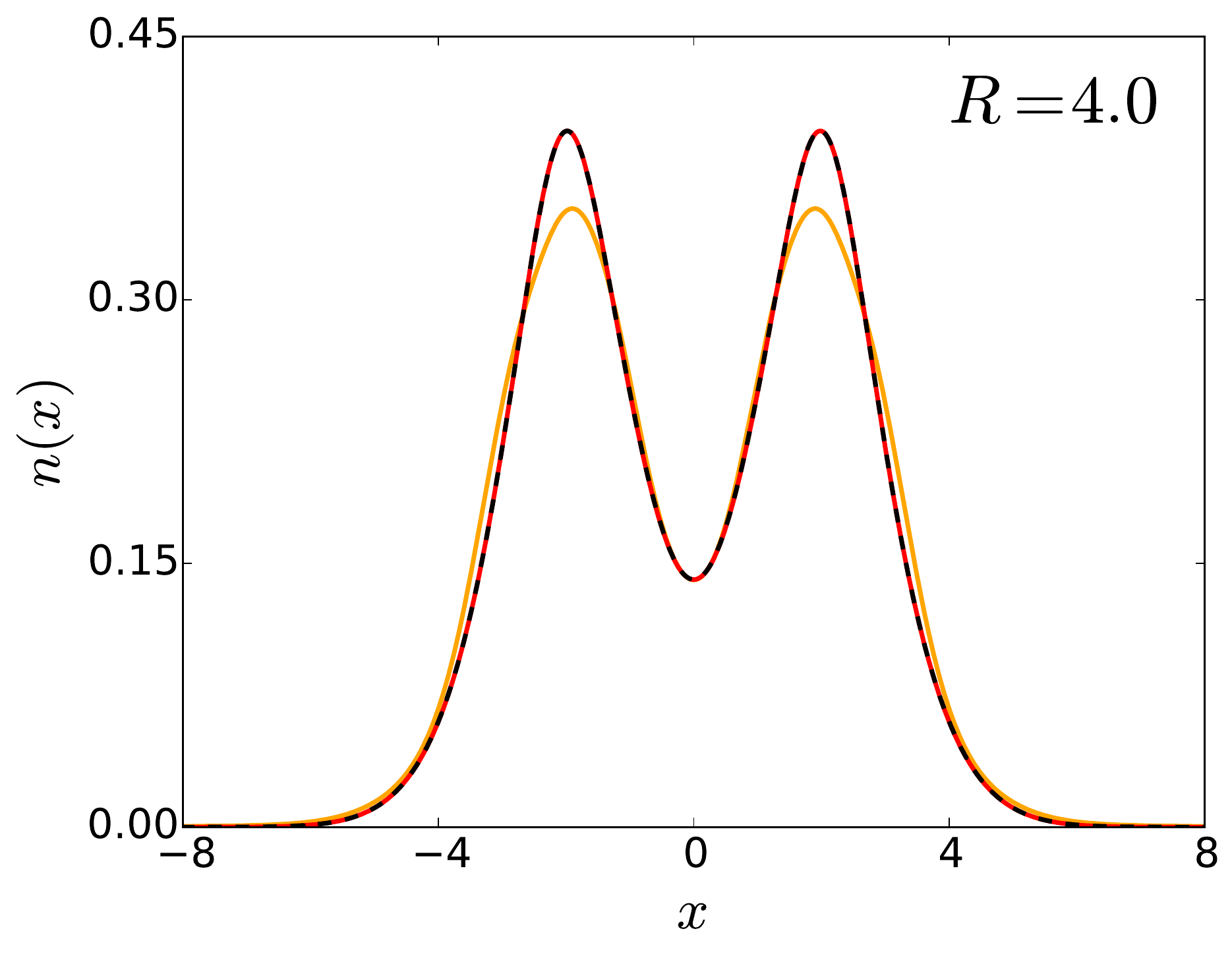}
\caption{(Color online) Optimal densities for 1d H$_2$ molecule in the
test set: DMRG (black dashed), ML with $N_T$=5 (orange solid), ML with $N_T$=20
(red solid).}
~
\vskip -0.8cm
\label{h2density}
\end{figure}
The problem is actually much greater than even the smooth dashed green curve would suggest.
In practice, we not only need the energy functional, but also its derivative, at least in
the vicinity of a solution density.   This is because we use the functional to find the
density itself, via
the Euler equation
\ben
\frac{\delta F}{\delta \n(x)}= -v(x).
\een
In fact, the derivatives of ML functionals such as that of Eq (6) are highly inaccurate
and cannot be used to find the minimizing density.  Methods have been developed
to constrain the search to the manifold of training data
via non-linear gradient denoising (NLGD) \cite{SRMB15}.
For our H$_2$ with $\NT=5$, these lead to the (even worse) solid green curve
of \Figref{h2a}.  The optimal density is shown in \Figref{h2density}.
We clearly see that (a) the accuracy is not high enough and (b) the 
error is dominated by the error in the densities.  (This is called a density-driven error \cite{KSB13}
in a DFT calculation.)  

However, when we increase to 20 data points, the ML curve (red solid)
is indistinguishable from the exact one, and the error at equilibrium is only 0.007 kcal/mol,
and shrinks with increasing $R$.  This calculation applies all the principles discussed
in Ref.~\onlinecite{SRHB13}, but is now applying them to the many-body problem, not just the KS problem.
Even in the stretched limit, where the
system is strongly correlated, there is no loss of accuracy.
Note that we are not just fitting the binding curve, as we are reproducing the many-body density
at every value of $R$, starting from data at a limited number of values.
In \Figref{h2density}, we plot the optimally-constrained densities at $R=4.0$ (outside all training sets)
for $\NT=5$ and $\NT=20$, compared with the exact density.

Here, ML has entirely bypassed the difficulty of solving the many-fermion
problem.  The machine learns the characteristics of the solution without ever solving the
differential equation.  Moreover, the HK theorem is a statement of the minimal information
needed to characterize the ground-state of the system.  
In some ways, this ML approach is the purest embodiment of the
HK theorem.

\subsection{Finding a data-driven optimal basis for longer chains}

The cost of optimal gradient descent methods, evaluated on a spatial grid,
grows very rapidly with the number of grid points, and rapidly becomes 
unfeasible as the number of H atoms grows.  Thus a simpler representation
of the density is required.
To overcome those difficulties, we introduce a basis set. 
Inspired by the localized atomic bases used in most quantum chemical codes, 
we developed a data-driven basis set using
Hirshfeld partitioning \cite{H77} and principal component analysis (PCA).

\begin{figure}[htb]
\includegraphics[width=0.45\textwidth]{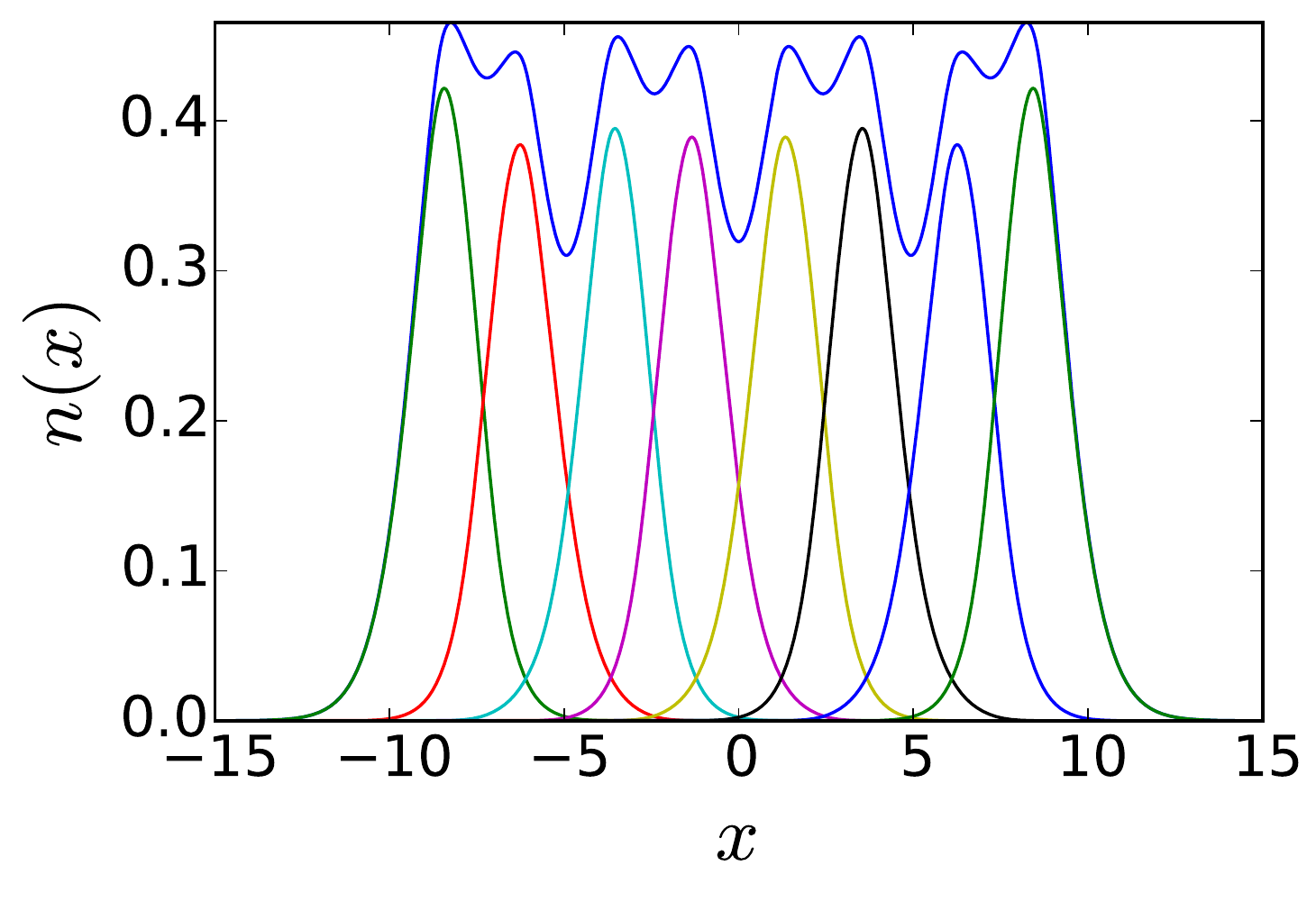}
\caption{Partition density of each H atom in H$_8$.}
\label{f:h77pcabasis_a}
\end{figure}
\begin{figure}[htb]
\includegraphics[width=0.45\textwidth]{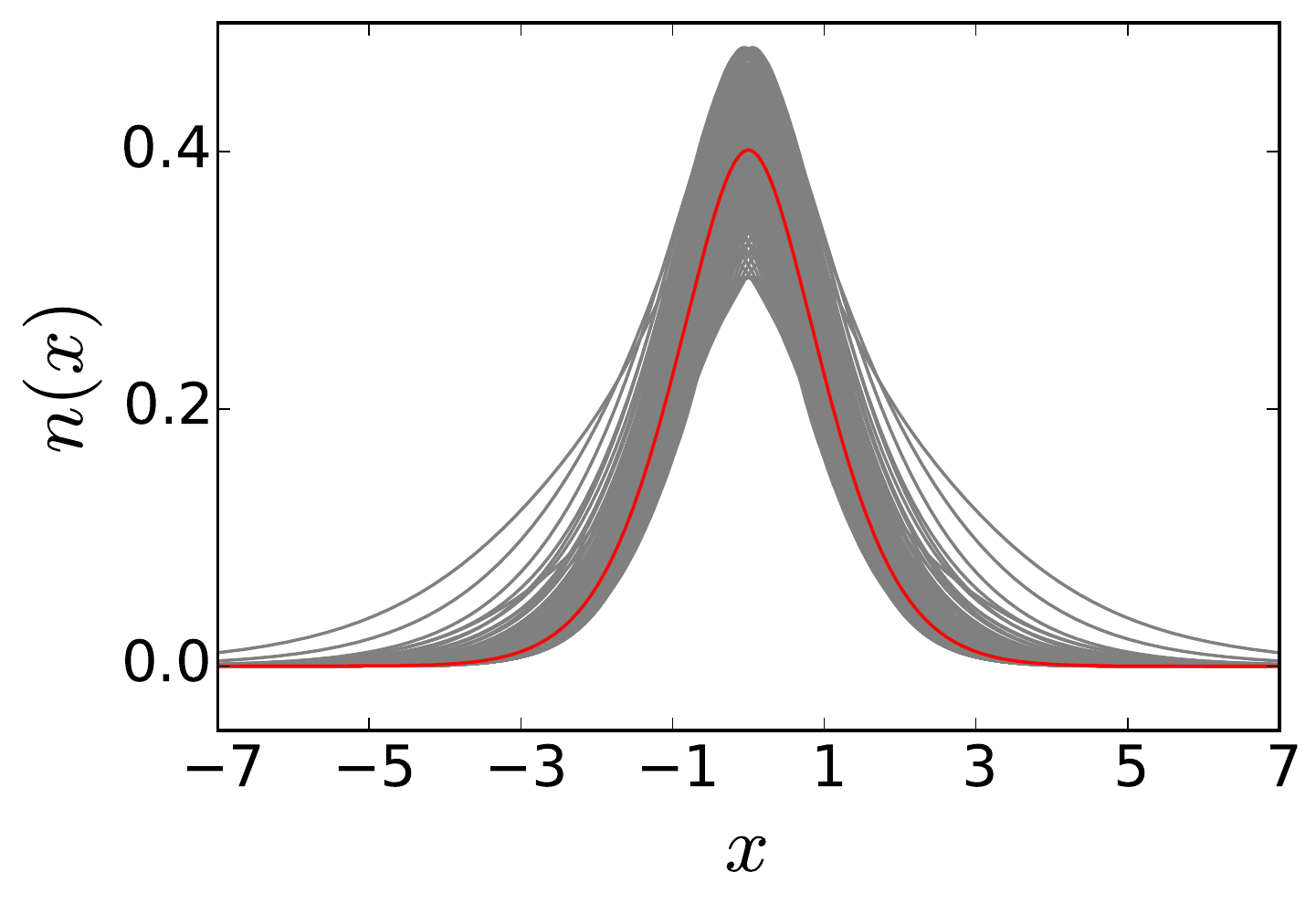}
\caption{
Single H atom densities for H atoms in different chains and atomic distance (gray).
The average density is plotted in red. }
\label{f:h77pcabasis_b}
\end{figure}
\begin{figure}[htb]
\includegraphics[width=0.45\textwidth]{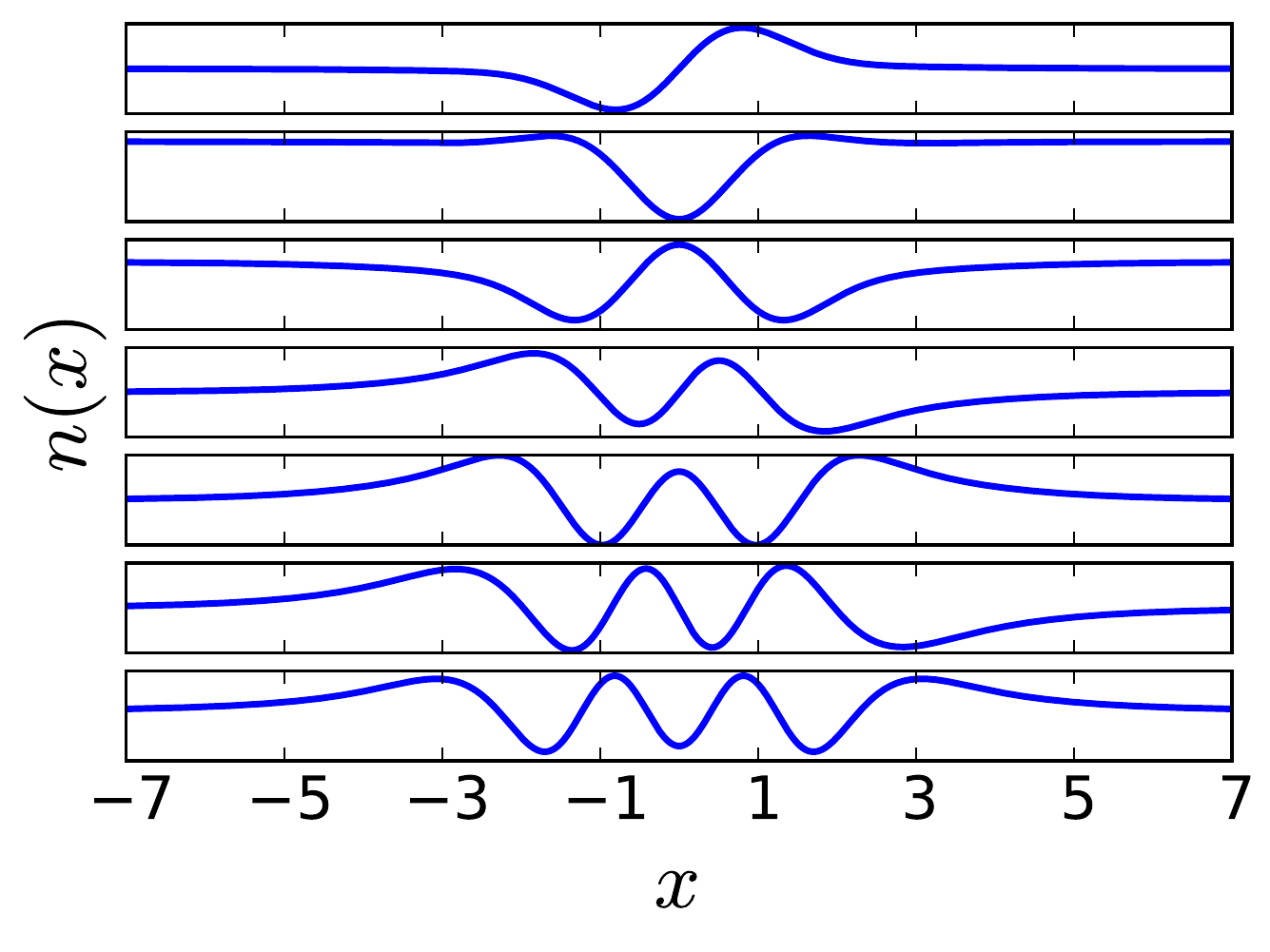}
\caption{
First 7 principal components of the densities shown in 
Fig. \ref{f:h77pcabasis_b}, from top to bottom.}
\label{f:h77pcabasis_c}
\end{figure}

To partition a molecular density via the Hirshfeld scheme, begin with
the protomolecule of overlapped atomic densities at the nuclear
positions of the real molecule.  If $\n^0_i(x)=\n^0_1(x-(i-1)R)$ is an isolated
atomic density at the $i$-th nuclear center, spaced $R$ apart, then
\ben
\n^0(x)=\sum_{i=1}^N n^0_i(x)
\een
is the  density of the protomolecule,
where $R$ is the interatomic spacing.  We define a weight
\ben
w_i(x)=\n^0_i(x)/\n^0(x),
\een
associated with each atom, and then define the density of each Hirshfeld atom within
the real molecule  as
\ben
n_i(x) = w_i(x) n(x),
\een
where $\n(x)$ is the exact molecular density.
The ground state density of a single hydrogen atom $n^0_i(x)$ is reported in
Ref.~\onlinecite{BSWB15}.
\Figref{f:h77pcabasis_a} shows partition densities $n_i(x)$ of atoms in one H$_8$.

Next, for a specific chain length $N$, we consider a range of interatomic separations
$R$, and consider the collection of every atomic density within the chain for every value of $R$
in a training set, each centered on the origin, as shown in   
\Figref{f:h77pcabasis_b}.
These individual atomic partition densities reflect the diverse
behaviors caused by the interaction between the hydrogen atoms inside the chains. 
A principal component analysis is applied to these densities, and the eigenvalues
are ordered in decreasing magnitude 
to find a subspace with the maximum variance.
Each atomic density
can be accurately represented by the base density $f_0(x)$
(red in \Figref{f:h77pcabasis_b}) and
7 principal components (\Figref{f:h77pcabasis_c}),
\ben
n_i (R,x) = f_0(x) + \sum_{p=1}^{7} c_{i,p}(R) f_p(x).
\een
Thus the total density of H$_N$ with separations $R$ is
$\sum_i^N n_i(R,x)$, and is described by just $7N$ coefficients. 
Note that $f_0(x)$ is very close to an isolated atom density, but
we use the average to center our data for the PCA analysis.
Our representation greatly reduces the number of variables in the
density representation for a given chain length, and saves a significant
amount of computational cost when solving for the ground state density of the
system. This new basis set is completely data-driven and physically meaningful.

\begin{figure}[htb]
\includegraphics[width=0.8\columnwidth]{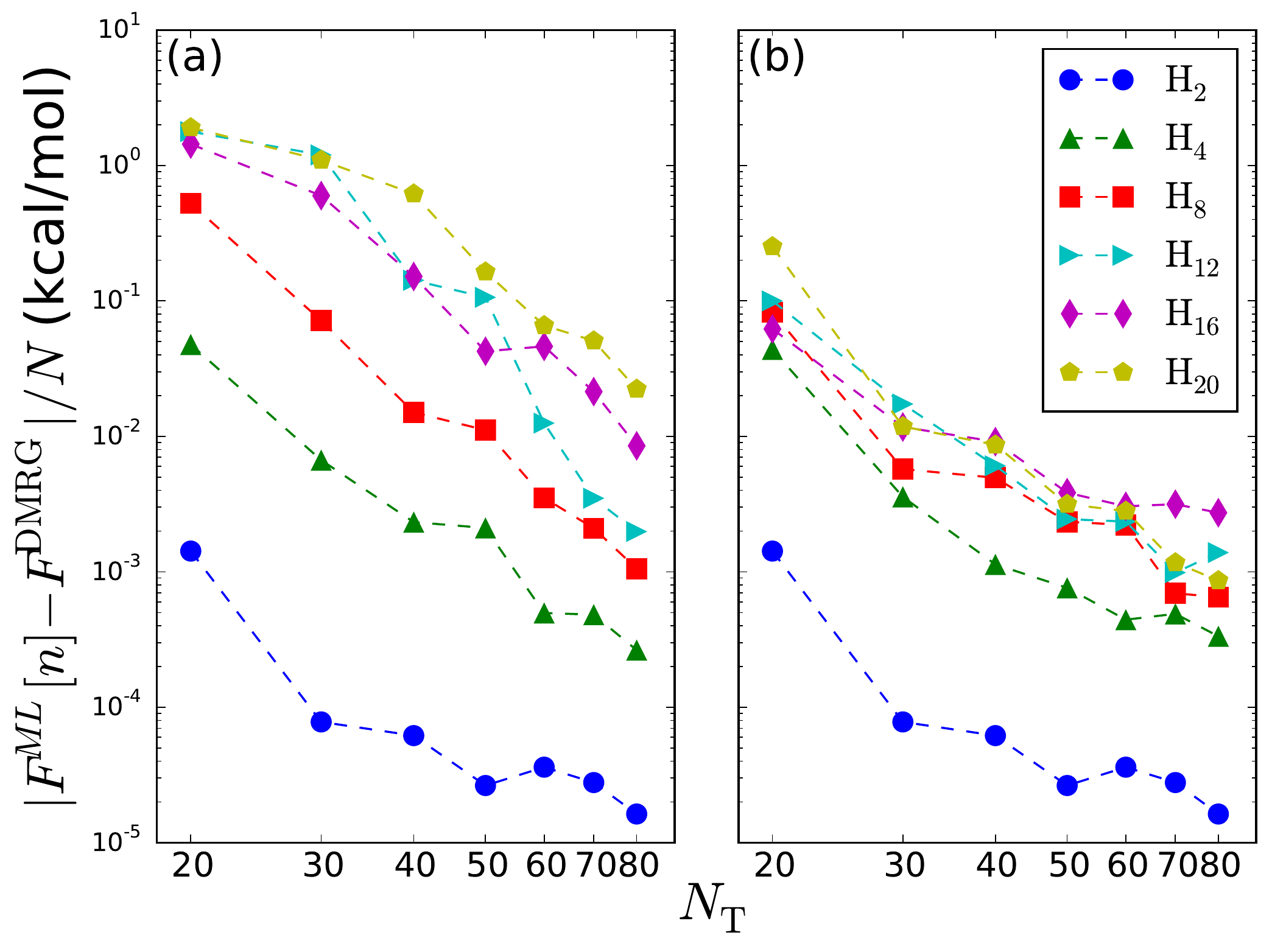}
\caption{(Color online) Learning curves for several 1d H chains.
(a)  ML using the total density. 
(b)  ML using the bulk partition densities (see text).}
~
\vskip -0.8cm
\label{lc_finite}
\end{figure}

\renewcommand*\arraystretch{1.4}
\begin{table*}[t]\footnotesize
\centering
\begin {tabular}{|c|c|c|c|c|c|c|c|c|c|}
\hline
$N$ & $N_\mathrm{T}$ & $\lambda$ & $\sigma$ & $\overline{|\Delta E_\mathrm{F}|} / N$ & $\max|\Delta E_\mathrm{F}| / N$ & $\overline{|\Delta E|} / N$ & $\max|\Delta E| / N$ & $E_{R=9.8}^\mathrm{ML} / N$ & $E_{R=9.8}^\mathrm{DMRG} / N$\\
\hline
2 & 5 & $1.0\times 10^{-8}$ & 1000 & 2.54 & 7.02 & 9.74 & 20.3 & -421.291 & -425.797\\
2 & 20 & $4.6\times 10^{-10}$ & 2.15 & 0.00121 & 0.00802 & 0.005 & 0.013 & -425.785 & -425.797\\
2 & 50 & $1.0\times 10^{-12}$ & 0.70 & 0.00003 & 0.00034 & 0.050 & 0.304 & -425.798 & -425.797\\
\hline
4 & 50 & $2.2\times 10^{-11}$ & 46.4 & 0.0021 & 0.016 & 0.005 & 0.017 & -428.617 & -428.620\\
8 & 50 & $1.0\times 10^{-4}$ & 2.15 & 0.011 & 0.31 & 0.28 & 1.68 & -430.011 & -430.032 \\
12 & 50 & $1.0\times 10^{-12}$ & 0.46 & 0.0031 & 0.010 & 0.24 & 0.88 & -430.502 & -430.503\\
16 & 50 & $2.2\times 10^{-11}$ & 0.46 & 0.0042 & 0.012 & 0.08 & 0.41 & -430.738 & -430.738\\
20 & 50 & $2.2\times 10^{-11}$ & 0.46 & 0.0042 & 0.014 & 0.26 & 0.88 & -430.880 & -430.880\\
$\infty$ & 50 & $1.0\times 10^{-8}$ & 0.46 & 0.012 & 0.050 & 0.073 & 0.27 & -431.447 & -431.444\\
\hline
\end {tabular}%
\caption{ML performance on different chains H$_N$. $N_T$ is the size
of training set. Regularization strength $\lambda$ and kernel length
scale $\sigma$ is the model hyperparameters selected by cross
validation \cite{LSPH14}.
The functional driven error $\Delta E_F/N$ \cite{KSB13} is tested on the entire
test set to show the overall accuracy. The total error $\Delta E/N$ is
tested on the
equilibrium test set to emphasize accuracy around equilibrium
position. $E_{R=9.8 / N}$ shows that ML can get very accurate
dissociation limit.
All errors are given in kcal/mol.
}\label{t:results}
\end{table*}

We next repeated these calculations for a sequence of chains of increasing length.
In each case, we train $F\ML[\n]$ on a limited training set, and then compare
on a test set (see supplementary material), with the accurate results supplied by DMRG.  
The learning curves, i.e., error as a function of $\NT$, of chains of differing
length, are shown in \Figref{lc_finite}(a).  The error typically decreases with 
amount of training data, but for fixed $\NT$, longer chains display substantially
larger errors.  This is because the binding energy curve changes more rapidly
when the chain length is increased.

In applied machine learning, feature engineering, which uses domain knowledge of
the data to improve the efficiency of ML algorithms, is a crucial step.
Here, we know that as the chain length increases, the central density should converge
to a fixed value (thermodynamic limit).  We therefore
choose the central two atomic densities alone to use
as a minimal input feature for learning the energy of a given finite chain.  
The learning curves for models trained only on this central partition density are
shown in \Figref{lc_finite}(b).  For chain lengths greater than or equal to
12, substantially greater accuracy is reached for a fixed amount of training data.
Here we still use the total density for $N\leq 8$ and the bulk density
for $N\geq 12$. The model performance and hyperparameters are presented in \Tabref{t:results}.

\subsection{Extrapolation to the thermodynamic limit}

Our ultimate goal is to use ML to find the energy of the infinite chain to within chemical
accuracy, for all interatomic separations.  
To do this, we first build
a set of infinite chain energies and densities.  For each value of $R$, we 
extrapolate both the density and energy of our finite chains as a function
of $N$.  This 
then gives us a set of data for the infinite chain that we can both train and test on
and gave rise to \Figref{Einf}.  

In an entirely separate calculation, we
also performed DMRG directly for the infinite chain, using the method of McCulloch \cite{mcculloch2008infinite} 
for a four atom unit cell \cite{itensor}.
The system is initialized by solving the equivalent finite size
system with box edges at $R/2$.  As a part of the iDMRG algorithm \cite{mcculloch2008infinite}, a single unit cell is then inserted into the center of the finite system and 15 sweeps are performed over the inserted unit cell. 
The sequence is repeated--after adding another unit cell--until convergence.
We compare these energies with the extrapolated values,
finding agreement to within 1 kcal/mol for all values of $R$.  This agreement validates our extrapolation
procedure.  We find that, with 50 data points, the ML result, on the optimized density,
also agrees to within 1 kcal/mol.   Thus, armed with the 50-data-point machine learned
functional, one can self-consistently find the density and energy of the infinite chain to quantum chemical
accuracy.

\begin{figure}[htb]
\includegraphics[width=0.8\columnwidth]{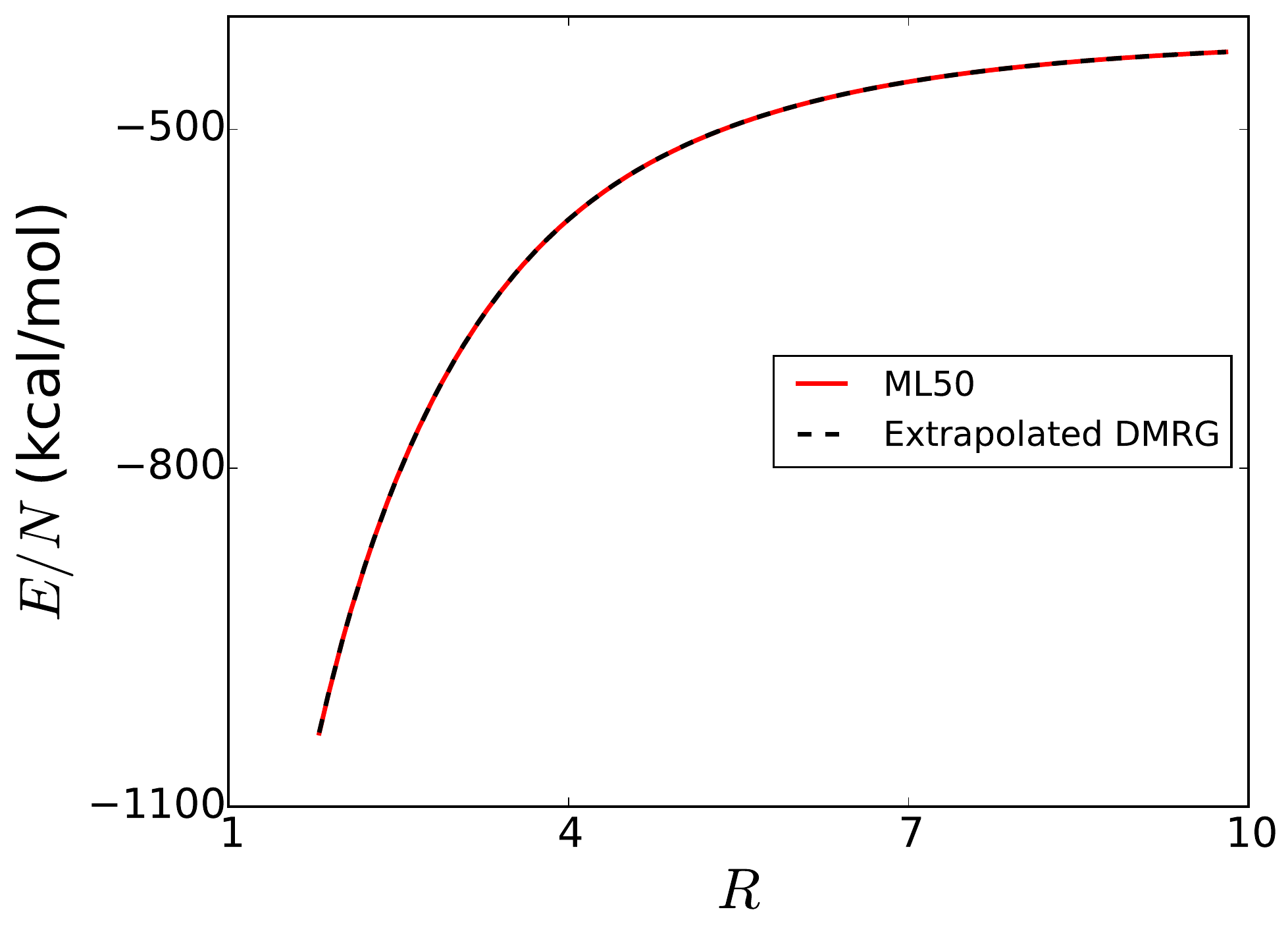}
\caption{(Color online) Electronic energy per atom in the thermodynamic
limit, both via DMRG chains (extrapolated to infinity) and using
machine learning with 50 data points per chain.}
~
\vskip -0.8cm
\label{lc_inf}
\end{figure}

\begin{figure}[tb]
\includegraphics[width=0.8\columnwidth]{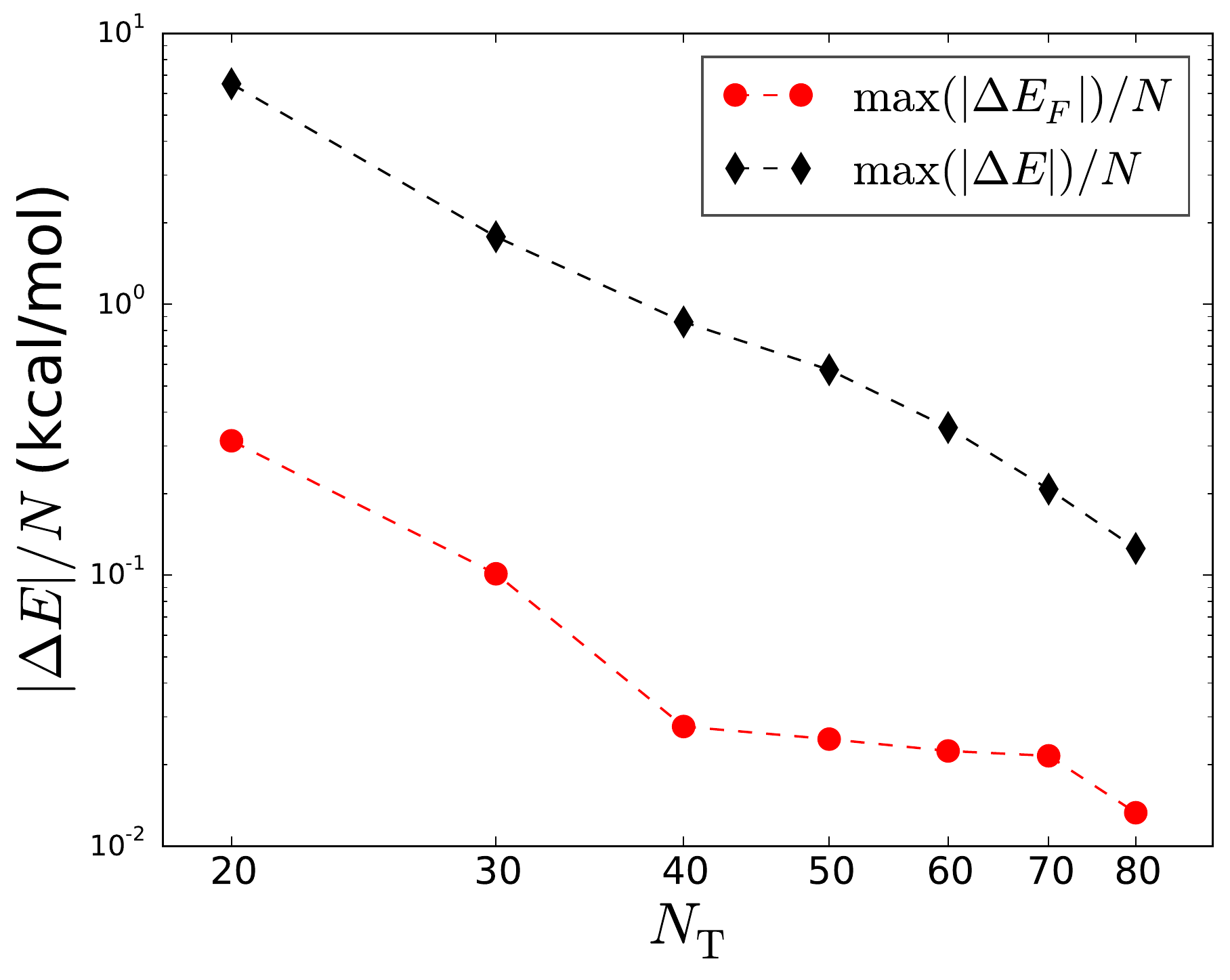}
\caption{(Color online) For a given training set with $N_T$ training points, the functional driven error, $\Delta F_F$ per atom is shown in red (lower curve).  The upper curve is the total energy error per atom evaluated self-consistently. \label{lc_finite}}
~
\vskip -0.8cm
\end{figure}
Our final figure simply demonstrates that the error for the infinite chain (and for all
the ML calculations) is almost entirely due to the error in the optimized density. 
The functional-driven error \cite{KSB13} is the energy error made on the exact
density:
\ben
\Delta E_F = E\ML[\n]-E[\n]=F\ML[\n]-F[\n].
\een
We see that, at any level of training, $\Delta E_F$
is an order of magnitude smaller than the final energy error on the
optimized density.  Thus the error is density-driven but, nonetheless, can be forced down
to quantum chemical limits with enough data.

\section{Discussion}

We have shown that it is in principle possible to construct, via machine
learning, the entire interacting functional of Hohenberg and Kohn, $F[\n]$,
so accurately that optimized densities and energies evaluated on them
are within quantum chemical accuracy.  
We have done this using the 1d simulation of continuum Hamiltonians
established over the last several years, and using DMRG as an efficient
solver.
We apply the ML methods previously developed for approximating the non-interacting
kinetic energy.  Here, because we have precise energies for the interacting
system to train on, we are able to construct the interacting functional, including
all exchange and correlation effects.
Our ML functional has no
difficulties when bonds are stretched so that correlations become strong.
We have even managed to apply this methodology to chains extrapolated to the
thermodynamic limit, producing chemically accurate  results for solids. 
This level of accuracy is far beyond that of any existing DFT calculation of a solid.

We conclude with a long list of the many things we have {\em not} done.
Most importantly, all calculations have been in one dimension, because of the
relative ease of setting them up, the efficiency of DMRG,  and the rapidity of approach to the bulk limit.
We do not know how much additional cost is involved in three dimensional calculations.
In fact, other recent work \cite{BLBM16}
shows that 3d molecular calculations do work for the
orbital-free problem, but that both a good choice of basis set is needed, as well as
building a density-potential map directly, avoiding the need to minimize the ML
density functional.  Thus, the extension to 3d appears entirely practical.

A second limitation is the rather large amount of data employed.  When we need an accurate
calculation at every 0.2 atomic units, how much have we achieved when we interpolate between
points?  But this extreme amount is needed only because we must produce a self-consistent
density from the ML functional, and one that is so accurate as to introduce
only a 1 kcal/mol density-driven error.  This is a level of accuracy that could only be
dreamed of with human orbital-free approximations.  There are two ways in which this requirement might
be very dramatically reduced:  (a) machine-learning the HK map $\n[v]$ directly, as mentioned
above (but here for the interacting density as a functional of the potential)
and (b) returning to a KS scheme.  The latter would yield highly accurate densities
almost always, and a ML $E\xc[\n]$ would almost certainly not produce a
significant density-driven error.  Of course, the price paid is the cost of solving the KS equations,
but that price is acceptable for the overwhelming applications of DFT at present.

A third important limitation is that, throughout this paper, we have used only uniformly
spaced chains of atoms.  In practice, of far greater interest would be the case of 
varying separations, even if all atoms (or molecules) are the same kind, as in an MD simulation
of liquid water.  We believe the density basis created here should prove very useful in
constructing a more general ML functional that would apply to a much larger variety
of situations.

In short, there are many issues that must be addressed before this functional could be
used in practice.  But in theory, machine-learning, combined with the
Hohenberg-Kohn theorem, can produce quantum chemical accuracy for strongly correlated solids.

\section{Acknowledgements}

This work was supported by the U.S. Department of Energy, Office of Science, 
Basic Energy Sciences under award \#DE-SC008696. 
T.E.B.~also thanks the gracious support of the Pat Beckman Memorial Scholarship
from the Orange County Chapter of the Achievement Rewards for College Scientists Foundation.

\clearpage
\section{Supplementary Material}
\subsection{Description of Data}
The density matrix renormalization group (DMRG) \cite{S91,S92,U05,U11} has become the gold standard for calculations in one dimension.  The ansatz made for the wavefunction is that of a matrix product state (MPS).  This ansatz allows for a site-by-site determination of the wavefunction by concentrating on a small number (in our implementation, two) lattice sites at a time.  Once the wavefunction is updated on those two sites, the next two sites are treated.  The entire system is swept back and forth until convergence which usually occurs very quickly in one dimension.

To evaluate the Hydrogen chains in this work, an extended Hubbard model \cite{SWWB12,WSBW12,BSWB15},
\begin{eqnarray}
\mathcal{H}&=&\sum_{j,\sigma}\frac{-1}{2a^2}\left(\hat c^\dagger_{j,\sigma}\hat c_{j+1,\sigma}+\mathrm{h.c.}\right)-\tilde\mu n_{j\sigma}\\
&&+\sum_j v^jn_j+\frac12\sum_{ij}v\ee^{ij}n_i(n_j-\delta_{ij}),
\end{eqnarray}
can be constructed to recover the continuum limit in the limit of many sites.  The prefactor on the kinetic energy terms is chosen to match the finite difference approximation for the kinetic energy with grid spacing $a$.  An external potential is applied in the variable $v^j$ while $\tilde\mu=\mu-\frac1{a^2}$ for chemical potential $\mu$.  Also, an electron-electron term, $v\ee^{ij}$ is represented by an exponential function \cite{BSWB15},  This exponential mimicks the soft-Coulomb interaction, which itself is an approximation of the Coulomb interaction in 3d but spherically averaged \cite{BSWB15}.  The similarity between these functions gives the similar behaviors of the 1d atoms and their 3d counterparts when the symmetry is high.

Systems are calculated with open boundary conditions (``box" boundary conditions).  The limit where the box boundary is far from the nearest atomic center is always taken, so no finite size effects appear.

A complication is apparent in 1d that does not appear in 3d.  There is no angular momentum in 1d.  Thus, not all neutral atoms bind their electrons.  One can see this in a reduced example as follows:  Consider a delta function interaction in 1d of the form $-\delta(x-R/2)-\delta(x+R/2)$ \cite{MB04}. When $R=0$, there is only one solution.  At any finite $R$, the number of electrons that will bind increases from two.  The same effect occurs for the exponential interaction, though it is not as easy to see.

This implies that a lower cutoff in the exponentially interaction hydrogen chains will impose a lower limit on suitable chain length.  We are interested in systems that do bind all electrons, so a systems below a critical $R$ are ignored. Table~\ref{tab:sup_data} lists the range of interatomic distances used for each chain.
For each Hydrogen chain data generated by DMRG, first sample 80 data from the \textit{entire test set range} in Table~\ref{tab:sup_data} equi-distantly. This test set is inaccessible in the training process. The rest of data in \textit{training set range} in in Table~\ref{tab:sup_data} are used as grand training set, where the $N_T$ training data are uniformly sampled to train the model. The \textit{equilibrium test set range} is a subset of entire test set range, emphasizing the performance around equilibrium positions. The upper bound is around twice the equilibrium position given by DMRG result.

\begin{table}[tb]
  \begin{center}
    \begin{tabular}{cccc}
      \toprule
      \hline
      $N$ & training set range & entire test set range & equilibrium test set range\\
      \hline
      \midrule
      $2$ & $1.0\leq R\leq 10$ (146) & $1.2\leq R\leq 9.8$ (80) & $1.2\leq R\leq 3.12$ (19)\\
      $4$ & $1.4\leq R\leq 10$ (136) & $1.6\leq R\leq 9.8$ (80) & $1.6\leq R\leq 4.08$ (25) \\
      $8$ & $1.4\leq R\leq 10$ (136) & $1.6\leq R\leq 9.8$ (80) & $1.6\leq R\leq 4.28$ (27)\\
      $12$ & $1.6\leq R\leq 10$ (131) & $1.8\leq R\leq 9.8$ (80) & $1.8\leq R\leq 4.32$ (26)\\
      $16$ & $1.6\leq R\leq 10$ (131) & $1.8\leq R\leq 9.8$ (80) & $1.8\leq R\leq 4.32$ (26)\\
      $20$ & $1.6\leq R\leq 10$ (131) & $1.8\leq R\leq 9.8$ (80) & $1.8\leq R\leq 4.4$ (27)\\
      \bottomrule
      \hline
    \end{tabular}
    \caption{Hydrogen chain data. $N$ is the number of Hydrogen atoms in the chain. $R$ is the atomic distance between atoms. The number of DMRG data in each range is in parenthese.}
    \label{tab:sup_data}
  \end{center}
\end{table}

\bibliography{refers,Master,adds}
\end{document}